\documentstyle[12pt]{article}
\def\L{{\cal L}}
\def\F{{\cal F}}
\def\wq{{$\bf w_Q(z)$}}
\input epsf
\begin{document}

\begin{center}
\Large{ Measuring the Equation-of-state \\
of the Universe: Pitfalls and Prospects
 }
 \end{center}

\begin{center}
Irit Maor$^{(1)}$, Ram Brustein$^{(1)}$,
Jeff McMahon$^{(2)}$,
Paul J. Steinhardt$^{(2)}$ 
\\
{\it $(1)$Department of Physics, Ben Gurion University, Beer-Sheva
84105, Israel \\
$(2)$ Department of Physics,
Princeton University, Princeton, NJ 08540 USA
}
\end{center}

\begin{abstract}

We explore  various pitfalls and challenges in determining the
equation-of-state ($w$) of dark energy component that dominates
the universe and causes the current accelerated expansion. We
demonstrated in an earlier paper the existence of a degeneracy
that makes it impossible to resolve well the value of $w$ or its
time-derivative with supernovae data. Here we   consider standard
practices, such as assuming  priors that $w$ is constant or
greater than  $-1$, and show that they also can lead to gross errors in
estimating the true equation-of-state. We further consider
combining measurements of the cosmic microwave background
anisotropy and the Alcock-Paczynski test
with supernovae data and find that the improvement in
resolving the  time-derivative of $w$ is marginal, 
although the combination can constrain its present
value perhaps  to 20 percent uncertainty.
\end{abstract}

\noindent
PACS number(s):  98.62.Py, 98.80.Es, 98.80.-k
\newpage

\section{Introduction}

Measurements of Type IA supernovae have shown that the expansion
of the universe is accelerating \cite{SCP,HZS}, suggesting that
most of the energy density of the universe consists of some form
of dark energy with negative pressure \cite{EOS}. Combining
measurements of the cosmic microwave background anisotropy and
observations of large-scale structure provides important
corroborating evidence.\cite{OS,Bahcall} Two candidates for the
dark energy are a cosmological constant (or vacuum density) and
quintessence,\cite{CDS} a time-varying, spatially inhomogeneous
component. In a previous paper \cite{Maor:2001jy} (Paper I), we
addressed the question of whether supernova measurements can be
used to measure the equation of state (EOS) of the negative
pressure component, the ratio $w$ of the pressure  to the energy
density. The issue is important because $w=-1$ for a cosmological
constant whereas $w$ takes on different values and can be
significantly time-varying in the case of
quintessence.\cite{CDS,other} Under the assumption that $w$ is
constant, its value can be determined to better than 5 per cent
by measuring several thousand supernovae distributed equally
between red shift $z=0$ and $z=2$. However, we showed that a
degeneracy opens up if $w$ is time-dependent which makes it
impossible to determine accurately the current value of $w$ or its
time-derivative. The cause of the degeneracy is that supernovae
measure luminosity distance, which is related by a multi-integral
expression to the EOS as a function of red shift, $w(z)$.  Widely
different $w(z)$ can have the same multi-integral value.

The purpose of this paper is to explore some pitfalls  and
challenges in determining the EOS and its time variation using
supernovae.  For example, we shall show how the standard practice
of  considering only models with $w\ge -1$ or only models with
constant $w$ when doing likelihood analyses can lead to grossly
incorrect results. For example, we will illustrate cases where
the standard practice will  suggest that  $w$ is near -1  or much
more negative than $-1$ when, in fact, $w$ is significantly
greater than -1 and rapidly time varying. We shall show that a
non-zero value of $dw/dz$ is more easily detected if $dw/dz>0$
than if $dw /dz <0$. We shall also contrast measuring $w$ for the
negative pressure component alone ($w_Q$)  versus the mean value
for the total energy density (including ordinary and dark
matter), $w_T$. The degeneracy  problem is less severe for $w_T$,
but this parameter provides less useful information. We consider
possibilities of breaking the degeneracy between $w$ and $d w/dz$
by combining supernovae results with either cosmic microwave
background anisotropy measurements and/or the Alcock-Paczynski
test. We shall show that neither additional test significantly
improves the measurement of the time-variation of $w$, although
optimistic assumptions about the Alcock-Paczynski test 
suggest that the current value of $w$ can be measured to within
20 percent or so.


We conclude that a new, yet to be found test has to be devised to
resolve well the cosmic EOS and its time variation. We stress
that with current data it is possible to determine the EOS to
about a factor of two. For a future experiment to significantly
enhance the determination of the EOS, and enable  the distinction
between a 
constant EOS and a time-dependent one, it needs to resolve 
the equation of state 
at the 10\% level or better.


The results of our analysis agree with many other analyses
\cite{Podariu,astier,Weller,Chiba,Barger,Chevallier,Goliath,trentham,gudmun,weller2,Ng,Kujat}, 
although not always with their interpretation, and can be used to explain
why some other analyses seem to indicate a superior resolving power of SN
measurements alone \cite{Hut,Staro,Boisseau,Wang}, or in combination with
other measurements \cite{Huterer,Tegmark,Corasaniti,Lovelace}. Some of the
latter analyses implicitly assume unrealistic accuracy in independent
determination of cosmological parameters, by not including
self-consistently the uncertainty in $w(z)$ in all measurements. For
example, assume a reported resolution of matter energy density which was
based on assuming $w_Q=-1$.

Type IA supernovae have intrinsic variability of about 0.15 in absolute
magnitude, but currently the errors in measuring distant SN are above
this. There are ongoing programs to extend the search to deeper
redshifts and improve measurement quality. The proposed SNAP satellite
plans to measure 2000 SN per year, mostly in the range of redshifts
$0.1<z<1.2$, and some as far as $z=1.7$.\cite{SNAP} The anticipated
error in individual magnitudes is $\Delta m=0.15$ statistical and 0.02
systematic error, which yields about 1\% relative error in luminosity
distance $d_L$.

For our numerical estimates, we have generated 50 SN magnitudes
randomly chosen from a uniform distribution in $z$ values,
between $z=0.1$ and $z=2$. Magnitudes were generated from a
Gaussian distribution with mean value $m(z)$  calculated using
fiducial models.  We have used our 50 points to simulate
approximately 2000 SN by reducing their magnitude error by a
factor $\sqrt{40}$ to 0.03 from the minimal 0.15 magnitude.  Thus
each generated point corresponds to 40 SNAP-like points, binned
together. This corresponds to $1.4\%$ relative error in $d_L$.
Hence, our analysis is based on
a SN search more extensive than the actual SNAP proposal.
To obtain a quantitative estimate of how well models are
resolved, we use one of two procedures. First,  we  can find the
maximum likelihood contours of the various models for each of the
fiducial models and explore the degeneracy in parameter space.
Alternatively, we can assume that all models which predict
$d_L(z)$ within 1\% of the fiducial cosmological model for all
$z$  between 0 and 2 are deemed indistinguishable. We find
that both approaches give comparable results.
That is,
the $95\%$ CL likelihood contours using the first procedure
are roughly equivalent to the indistinguishability region of the
second.

\section{Dependence of luminosity distance on dark and total EOS}
\label{dl}

Luminosity distance is defined to be the ratio of luminosity $\L$
to flux $\F$
\begin{equation}
    d_L=\sqrt{\frac{\L}{4\pi\F}}=(1+z)r,
\end{equation}
where  the present value of the scale factor $a_0$ is
normalized to unity
(throughout
 subscript 0 denotes present values);
 $r$ is the coordinate
 distance
\begin{equation}
    r=\int_1^{1+z}\frac{dx}{H};
\label{cd}
\end{equation}
and
$H$ is the Hubble parameter $H=\dot{a}/a$. The observed SN
magnitudes are related to $d_L$,
\begin{equation}
m(z)=M+25+5\log_{10} \left[ H_0 d_L(z) \right] ,
\end{equation}
$M$ being the SN absolute magnitude.

For a flat universe with two energy sources,  matter (including
dark matter) and a dark Q-component, there are two equivalent
routes to computing $d_L$ without assuming any prior about the
time dependence of $w_Q$. One way is to use the algebraic relation
between the total energy density and the Hubble parameter $H$.
Using conservation equations, the energy densities of the dark
component $\rho_Q$ and that of ordinary matter $\rho_m$, are
given by
\begin{eqnarray}
 \label{four}
  \rho_Q (z)&=& (\rho_Q)_0(1+z)^3\  exp\left[3\int_1^{1+z}
w_Q\frac{dx}{x}
    \right]  \\
   \rho_m(z) &=&(\rho_m)_0(1+z)^3.
\end{eqnarray}
Since
\begin{eqnarray}
\left(\frac{H}{H_0}\right)^2&=&\frac{\rho_m+\rho_Q}{(\rho_m)_0+(\rho_Q)_0}
 \nonumber \\ &=&
     \frac{g}{1+g}(1+z)^3+\frac{(1+z)^3}{1+g}\ exp\left[
     3\int_1^{1+z}w_Q\frac{dx}{x} \right]
\end{eqnarray}
where $g$ denotes the present ratio of matter to dark energy
densities $g=(\frac{\Omega_m}{\Omega_Q})_0$, $H$ can be expressed
as
\begin{equation}
  H=H_0\
(1+z)^{3/2}\left(\frac{g}{1+g}+\frac{1}{1+g} exp\left[3\int_1^{1+z}
    w_Q\frac{dx}{x} \right] \right)^{1/2}.   \label{ein}
\end{equation}
Substituting this into $d_L$ gives
\begin{eqnarray}
   d_L
&=& \frac{(1+z)}{H_0}(1+g)^{1/2}\int\limits_1^{1+z}
    \frac{d x}{x^{3/2}}
    \left[g+exp\left(3\int\limits_1^{x} w_Q(y)\frac{dy}{y}
    \right)\right]^{-1/2}
    \label{dl1}
\end{eqnarray}

An equivalent approach is to treat the sum of dark matter and
the dark energy component as a single cosmic fluid with
average equation of state
$w_T(z)=\Omega_Q(z)
w_Q(z)$, where
\begin{eqnarray}
   w_T(z)&=&\frac{w_Q}{1+(\rho_m/\rho_Q)}
\nonumber \\ &=&
    \frac{w_Q}{1+g\  exp\left[-3\int_1^{1+z}w_Q(x)\frac{dx}{x}.
    \right]}
\end{eqnarray}
Since $H^2$ is proportional to the total energy density in the
universe, it can be expressed in terms of $w_T$ as follows
\begin{equation}
 \label{seven}
  \left(\frac{H}{H_0}\right)^2
  =exp\left[ 3\int_1^{1+z}\left(1+w_T(x)\right)\frac{dx}{x}
  \right].
\end{equation}
Here we have used the conservation equation for the total energy
density. Using (\ref{seven}) we can express $d_L$ in terms of
$w_Q$ as follows,
\begin{eqnarray}
 \label{eight}
   d_L&=&\frac{(1+z)}{H_0}\int_1^{1+z} dx\ exp\left[-\frac{3}{2}
    \int\limits_1^{x}(1+w_T(y))\frac{dy}{y} \right]= \nonumber \\
  &=&\frac{(1+z)}{H_0}\int_1^{1+z}\frac{dx}{x^{3/2}}
exp\left[-\frac{3}{2}\int\limits_1^{x}\left(\frac{w_Q(y)}{1+g\
    exp\left[-3\int_1^{y}w_Q(u)\frac{du}{u} \right]}
    \right)\frac{dy}{y}\right].
\end{eqnarray}
This expression for $d_L$ as a function of $w_Q$ has
 one more integral than the relation in Eq.~(\ref{dl1}),
 but it is pedagogically useful in
 demonstrating that  $d_L$ is sensitive only
to a weighted average of $w_T$ or $w_Q$ and not to their detailed
time dependence.

\section{Constraining dark and total EOS using SN measurements}
\label{wqwt}

Based on the previous sections, a number of lessons can be
learned about measuring the EOS.  First,
the relation
between $w_T$ and $d_L$ in (\ref{eight}) involves an
integral, so we do expect some degeneracy in the determination of
$w_T(z)$ from SN measurements.
To determine the EOS of the dark energy itself, $w_Q$,
the total energy
density must be resolved
into a matter component and a dark energy component.  Hence,
if $\Omega_m$ and $\Omega_Q$ are not known
from independent measurements,
determining $w_Q$ entails an additional uncertainty.
For example, consider a flat universe with $\Omega_T =
\Omega_m + \Omega_Q =1$.
Since the matter EOS is $w_m=0$,
it follows that $w_T = w_Q \Omega_Q$.
Two models with different values of $w_Q$ may produce
the same value of $w_T$ and, consequently, $d_L$, due to offsetting
differences in the value of $\Omega_m$.

The more negative $w_T$ is, the faster is the expansion.  Therefore, a
more negative (positive) $w_T$ will make $d_L$ larger (smaller). In
addition, light from earlier times (emitted at higher $z$ values) must
pass through the low $z$ universe to reach us. This means that changes in
$w_T$ at lower $z$ affect $d_L$ at higher $z$.  The converse is not true. 
Changes in $w_T$ at high $z$ do not affect $d_L$ at lower $z$.

\begin{figure}
  \begin{center}
 \epsfxsize=3.3 in \centerline{\epsfbox{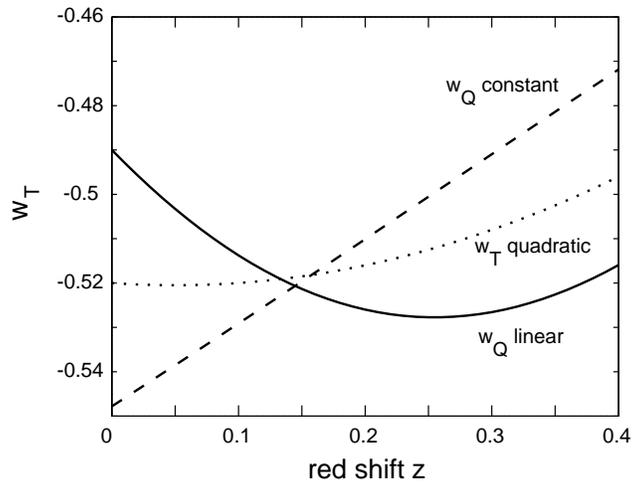}}
  \end{center}
  \caption{
   \label{qln8tot}
   $w_T (z)$ for three best fit models of three fits under three
   different assumptions: constant $w_Q$ (dashed), linear $w_Q$
   (solid), and quadratic $w_T$ (dotted),
    to data generated from a single fiducial model:
    $(w_Q,\Omega_m)=(-0.7-0.8z, \, .3)$.
    All fits prefer $w_T^*\equiv w_T(z^*\simeq .15)\simeq -.52$,
    but diverge for other values of $z$.}
\end{figure}

Consequently, it is not surprising that SN measurements of $d_L$
provide stronger constraints on $w_T(z)$ at low $z$ than at high
$z$. In particular, if all cosmic parameters other than $w_Q(z)$
are fixed, there is a particular, relatively low value of $z=z^*$
for which $w_T(z)$ is most tightly constrained. This value of
$z^*$  is clearly seen in our numerical results and was noted
independently by \cite{astier,Huterer}. For example,
Figure~\ref{qln8tot} shows the EOS $w_T(z)$ for three models each
of which is obtained by best fit to $d_L(z)$ for a fiducial model
$(w_Q,\Omega_m)=(-0.7-0.8z, \, .3)$ using one of three fitting
assumptions: (1) that the dark EOS is constant; (2) that the dark
EOS is linear $w_Q=w_0+w_1 z$; and,  (3) that the total EOS is
quadratic $w_T=A + B z + C z^2$. While the real degeneracy is
stronger than what is seen in the figure, we have chosen three
examples to illustrate the existence of $w^*_T$ and $z^*$. As can
be seen, the fits disagree significantly for $z$ far from
$z^*=0.15$, but all fits agree near $z^*$.

\begin{figure}
  \begin{center}
    \epsfxsize=3.3 in \centerline{\epsfbox{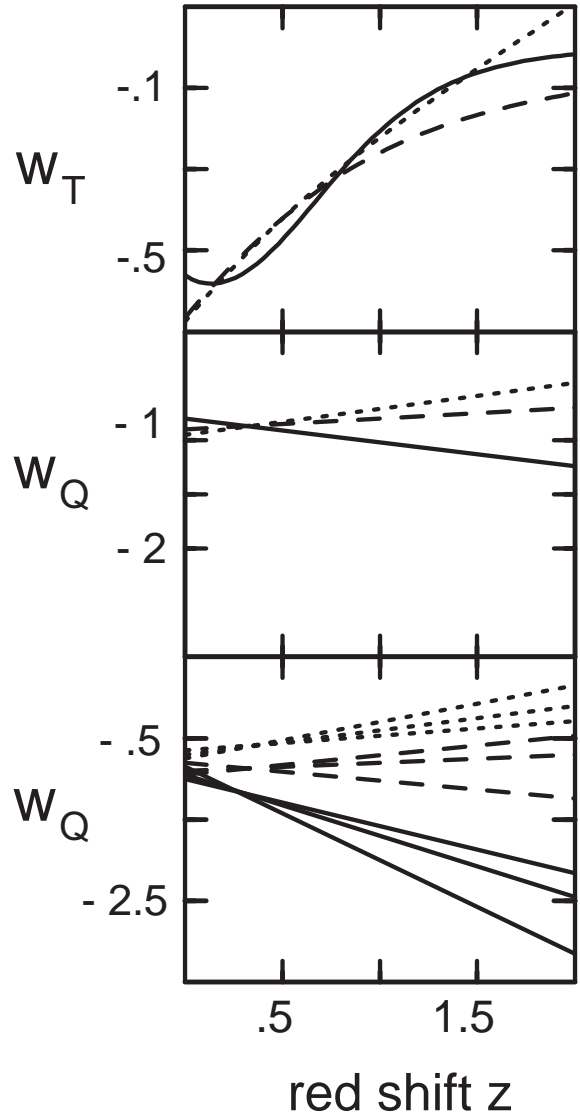}}
  \end{center}
  \caption{Models within 95\% CL region of a fit
   to data generated from the fiducial model $(w_Q,\Omega_m)=(-1,0.3)$
   assuming $w_Q=w_0+ w_1 z$.
  Top: The total EOS $w_T(z)$, for three different linear models.
       Middle: $w_Q(z)$ assuming $\Omega_m=0.3$ exactly, for the
same
       linear models.
       Bottom: $w_Q(z)$ for nine
       models, assuming that $0.2<\Omega_m<0.4$ (no relation
       between the dashed, dotted and solid lines of the
       bottom panel to those of the middle and top ones).
    \label{evolution}}
\end{figure}

Unfortunately, the resolution of $w_Q(z^*)$, the quantity which
most interests us, is  degraded when
we do not fix $\Omega_m$ but, instead, allow for the current
uncertainty in its value.
In Fig.~\ref{evolution}, we  show  some linear fits
to simulated data
generated from the fiducial model $(w_Q,\Omega_m)=(-1,.3)$.
The fits are representative examples  which
fit the fiducial model to within  the 95\% confidence region.
 The upper plot shows that $w_T(z)$ (with $\Omega_m$
fixed at 0.3) is relatively well resolved, and particularly well
resolved at around redshift $z^*=.3$. The resolution is not that
sharp in the middle plot which shows the corresponding $w_Q$, but
a special point of enhanced resolution around $z=.4$ is still
clearly seen. If one lets $\Omega_m$ vary in the realistic range
of $0.2-0.4$, then $w_Q$ becomes poorly resolved and the spread
at $z^*$ increases significantly. Similarly the spread in $w^*$
and $z^*$ increases significantly if more general functional
forms of the EOS are considered.

\begin{figure}
  \begin{center}
    \epsfxsize=3.3 in \centerline{\epsfbox{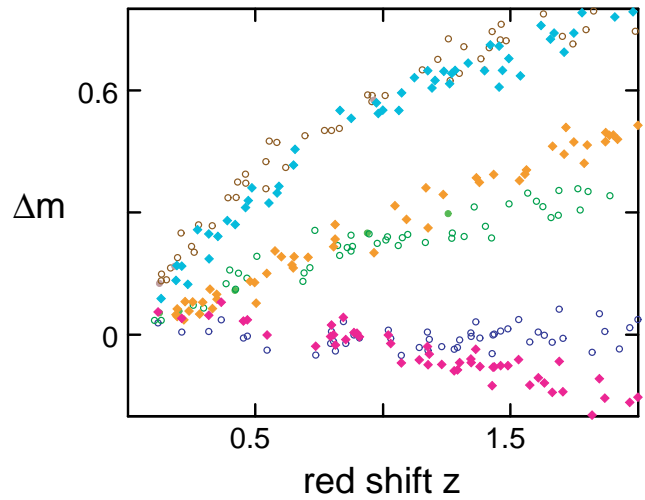}}
  \end{center}
  \caption{Magnitude differences between pairs of degenerate models
  and a flat pure matter ($\Omega_m=1$) Universe. Each pair consists of
  simulated data points generated from one
  constant $w_Q$ model (open circles) and one linear $w_Q$ model
  with a large (positive or negative)
  derivative (full squares). The pairs are well separated but it is hard
  to separate between ``members" of each pair.
    \label{sciam}}
\end{figure}

We would like to stress that the constant or linear forms of $w_Q$ that
we use are not meant to be anything more than simple concrete examples
to highlight the fact that we are dealing with a degenerate parameter
space.  Showing that if one assumes a linear $w_Q(z)$, then it can be
resolved to, say, 50\% does not logically mean that it can be measured
to 50\% accuracy generally since $w_Q(z)$ is resolved with different
accuracy depending on its functional form. This can be illustrated with
the following examples. In Fig.~\ref{sciam}, the difference in
magnitude ($\Delta m$) for models with various EOS is shown. There are
three clusters of points, each of which corresponds to a simulation of
SN data for pair of different models.  Each pair consists of a constant
and linear $w_Q$.  Each pair can be clearly separated from other pairs
but the constant and linear ``members" of a pair cannot be
distinguished by SN data.
The examples chosen for Fig.~\ref{sciam} have unrealistic large
derivatives (of order unity)  and therefore start to diverge from
their constant partners for large $z$. More realistic examples
with smaller derivatives or oscillatory behaviour will be much
harder to distinguish from a constant EOS.

Clearly the treatment of the SN analysis is important. If it is assumed
that $w_Q$ is constant, the figure shows that different values can be
resolved to high accuracy, but if the assumption of constancy is
relaxed and a linear $z$ dependence is allowed it becomes clear that
the data can determine well only a single relation between $w_0$ and
$w_1$ and that $w_Q(z)$ is poorly resolved.

\begin{figure}
  \begin{center}
    \vspace{-.1in}
    \centerline{
    \epsfxsize=75mm \epsfbox{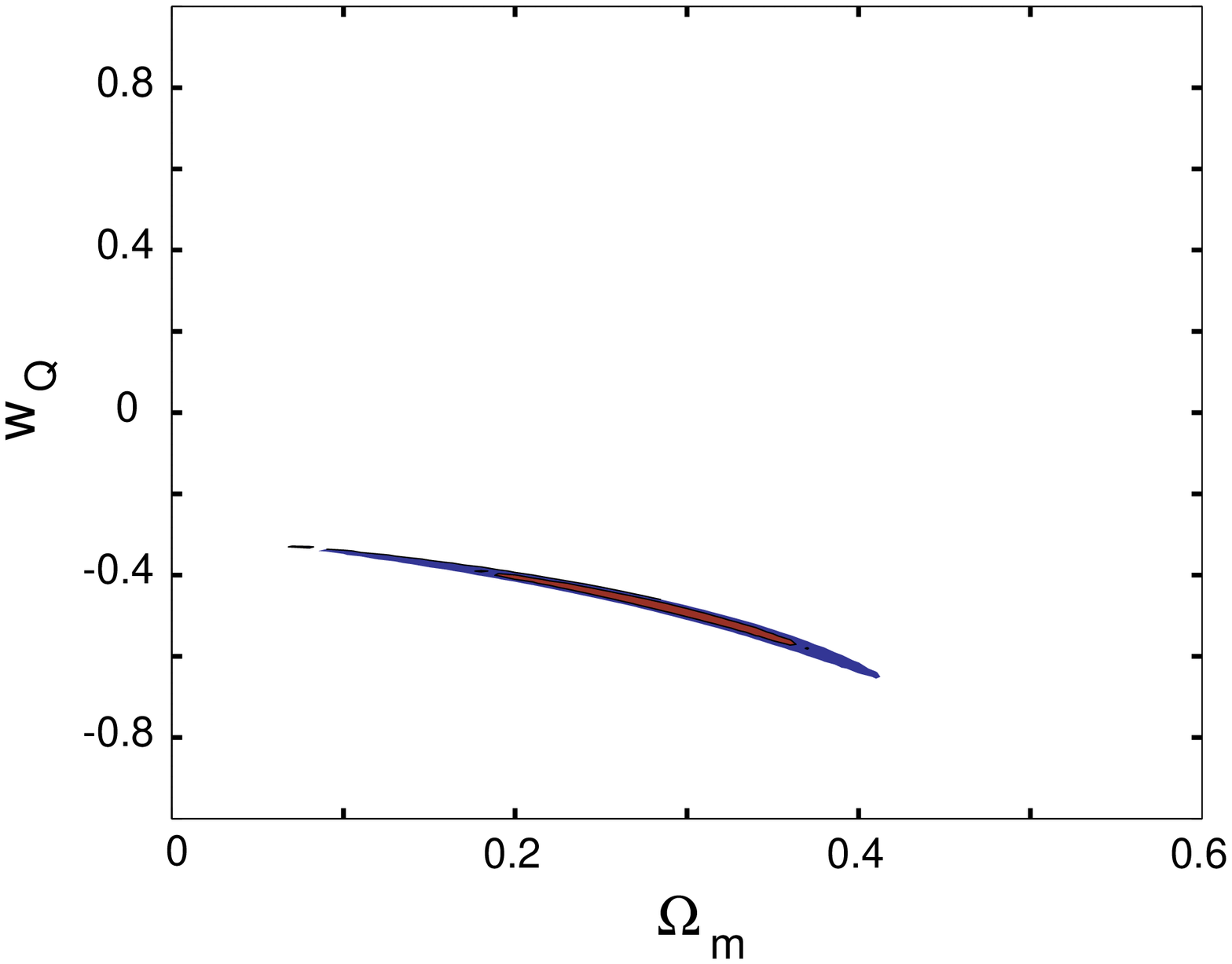}  
    \epsfxsize=75mm \epsfbox{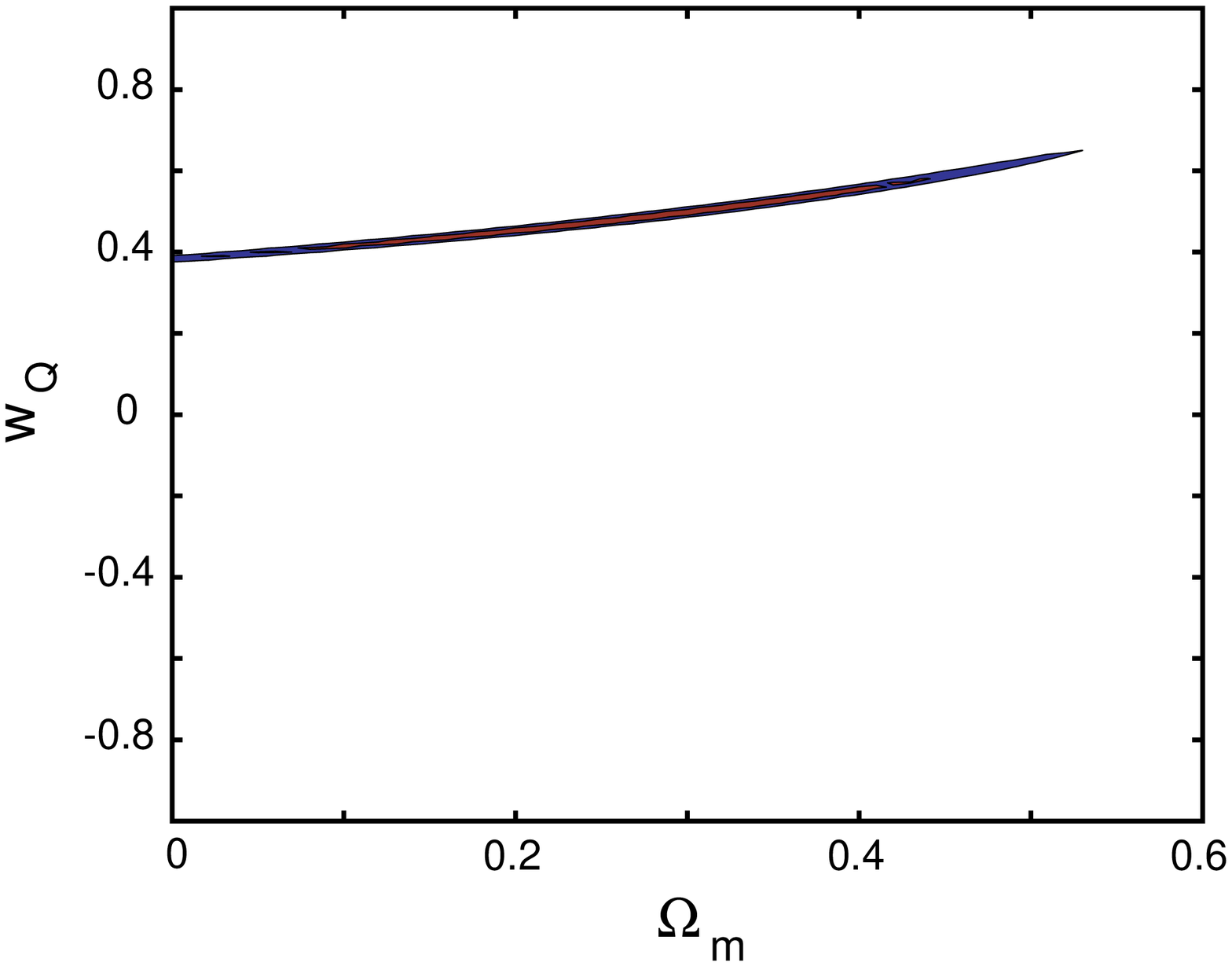}}
    \vspace{-.2in}
  \end{center}
  \caption{ 95\% CL contours of fits to data generated from
   two fiducial models.
  The curvature of  degeneracy contours is positive
  for a positive $w_Q$ fiducial model (right)
  and negative for a negative $w_Q$ fiducial model (left).
    \label{qc5qc}}
\end{figure}

The degree of degeneracy exhibited in the $w_Q=const.$ fits depends on
 whether $w_Q$ is positive or negative. Recall that if different
models yield a total EOS $w_T=w_Q \Omega_Q$ that is approximately
equal, they are degenerate, and therefore changes in $w_Q$ can be
compensated by changes in $\Omega_Q$ (or equivalently, in
$\Omega_m$). The difference between the case where $w_Q$ is
positive is due to the specific way in which this compensation
mechanism operates.  If $w_Q$ is positive, the curvature of
degeneracy lines in $(w_Q,\Omega_m)$ plane is  positive, as shown
 in the right panel of Fig.~\ref{qc5qc}
for a fiducial model with  $w_Q=+0.5$.
 Conversely, if $w_Q$ is negative,
the curvature of the degeneracy line in $(w_Q,\Omega_m)$ plane is
 negative,
as demonstrated in the left panel of figure~\ref{qc5qc}.
We have found that this result is unaffected by the
 value of the
derivative of $w_Q$, even if it is quiet large.

\vspace{0.5in}

\section{Common Practices and Pitfalls in Determining
\wq} \label{wqge-1}

The previous section (and Paper I) show that the determination of
$w_Q(z)$ from SN data is a more delicate process than it would
seem.  If we know {\it a priori} that $w_Q$ is constant, then
its value can be determined quite accurately.  However, without
this assumption, $w_Q$ is poorly determined, and matters much
get much worse if $\Omega_m$ is uncertain.

The analysis can be further confounded if certain common practices
are followed.  For example,
many analyses  assume that $w_Q$ is constant and presume that,
even if $w_Q$ is time-varying, the constant- $w_Q$ fit will
provide the mean value over recent epochs.
Another common practice is to
impose the condition that  $w_Q(z)$ be
limited to $-1 \le w_Q(z) \le 1$, based on the
positivity and stability conditions that apply to most (but not
all) forms of dark energy.
We shall see that both practices can produce enormous distortions
of the likelihood surface that lead to grossly incorrect conclusions.

\begin{figure}
  \begin{center}
    \vspace{-.5in}
     \epsfxsize=120mm \centerline{\epsfbox{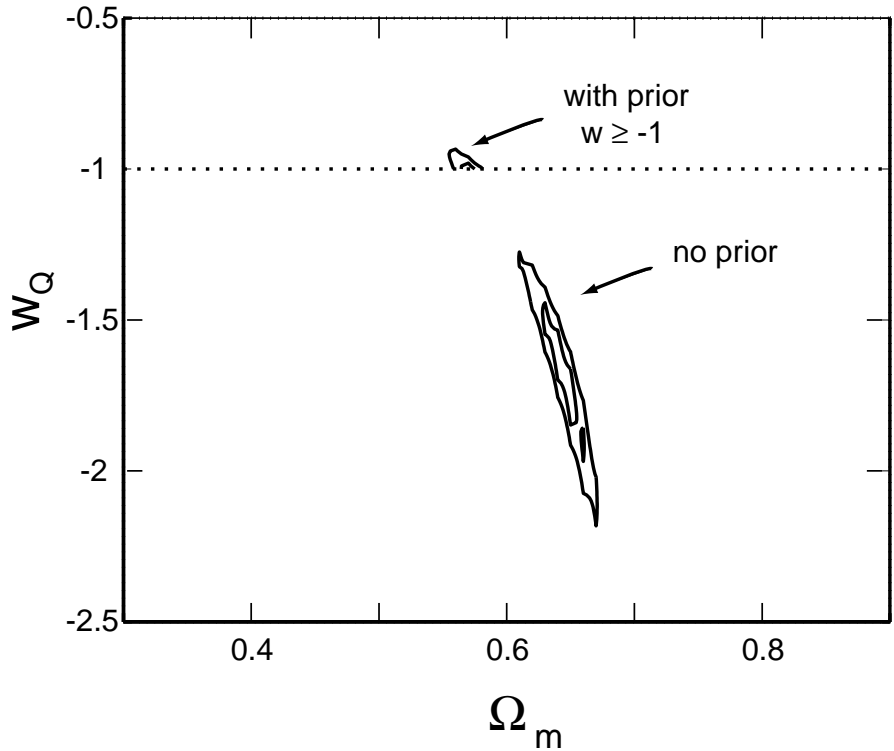}}
    
    \vspace{-.2in}
  \end{center}
  \caption{ Constrained (small) and unconstrained (larger and more
  negative) 68\% and 95\% confidence contours of a fit to data
  generated from a fiducial model with linear $w_Q$
  $(w_Q,\Omega_m)=(-.7+.8 z,.3)$. The fit is done under the
  (wrong) assumption that $w_Q$ is constant.  \label{cutqlp8}}
\end{figure}

For example, we have tried to fit
data generated from a fiducial model with $w_Q=-.7+.8 z$, and
$\Omega_m=.3$ over a redshift range $0<z<2$.
Note that the fiducial model has $w_Q>-1$ for all $z$.
Yet, if we do a best-fit assuming the prior that $w_Q$ is constant,
we find it to be
$(w_Q,\Omega_m)\simeq(-1.75,.65)$. Not only does the best-fit
have $w_Q<-1$, but  the whole 95\% confidence
contour lies in a region where $w_Q<-1$. The results are  the elongated
contours in the lower part of
Fig.~\ref{cutqlp8}. The reason for such a strange result can be
understood from the functional dependence of $d_L$ and $w_Q$.
Assuming that $w_Q(z)=w_0+ w_1 z$, the energy density of the dark
component is given in eq.(\ref{four}),
\begin{equation}
  \rho_Q=(\rho_Q)_0(1+z)^{3(w_0-w_1+1)}\, {\rm exp}[3w_1z],
\label{rhoq}
\end{equation}
so  increasing $w_1$ and decreasing $w_0$ have opposite and
compensating  effects, which tend to cancel each other's influence
on $d_L$. The exponential factor determines the quantitative
details of this compensation since changes in $w_0$ need to
compensate also for changes in the exponential factor. It is
therefore clear that the compensation cannot be perfect over a
range of $z$'s. If we insist on the prior of constant EOS ({\it
i.e.}, $w_1=0$), the fitting procedure will pick out a value of
$w_0$ which is much  more negative than the fiducial value. In the
figure we have picked a fiducial with a large positive derivative
to illustrate our point, but it is clear from our discussion that
the same problem arises  when time dependence is weaker, or in
cases that the EOS has a more general functional form.

Introducing a prior that $w_Q>-1$ can give a very misleading impression
of how well $w_0$ is resolved. For example, suppose that we assume the
priors that $w_Q$ is constant and $w_Q>-1$, as is standard practice.
The results are shown in Fig.~\ref{cutqlp8}, the small contours
truncated at $w_Q=-1$. They seem to suggest that the data supports the
conclusion that $w_Q=-1$ with a high level of confidence. The best fit
is $(w_Q,\Omega_m)\simeq(-1,.58)$. Yet, this is not related in any
obvious way to the fiducial model, $w_Q=-0.7 + 0.8z$ and
$\Omega_m=0.3$.
The values of $\chi^2$ per degree of freedom for the
best fit models of Fig.~\ref{cutqlp8} are reasonable, .95 for the
unconstrained fit, and 1.39 for the constrained fit.  So, what appears
to be a compelling result is actually a total distortion.  Of course,
it is also conceivable that the actual $w_Q(z)$ is less than $-1$, in
which case the same procedure of introducing a prior would falsely
suggest that $w=-1$ fits well.

\vspace{0.5in}

\section{Asymmetry in determination of  the time dependence
of the dark EOS}

\begin{figure}[h]
  \begin{center}
    \centerline{
    \epsfxsize=50mm \epsfbox{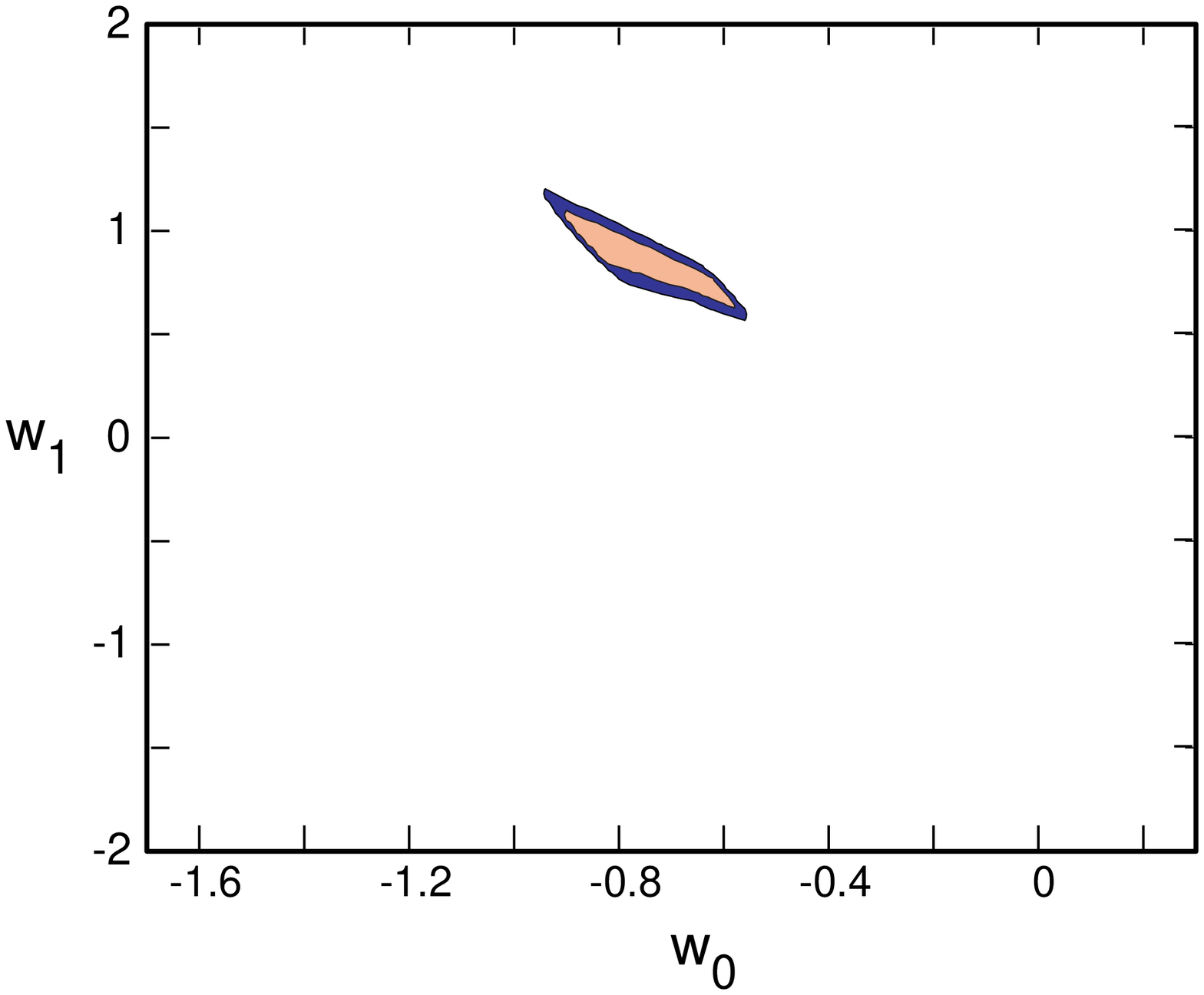}  \
    \epsfxsize=50mm \epsfbox{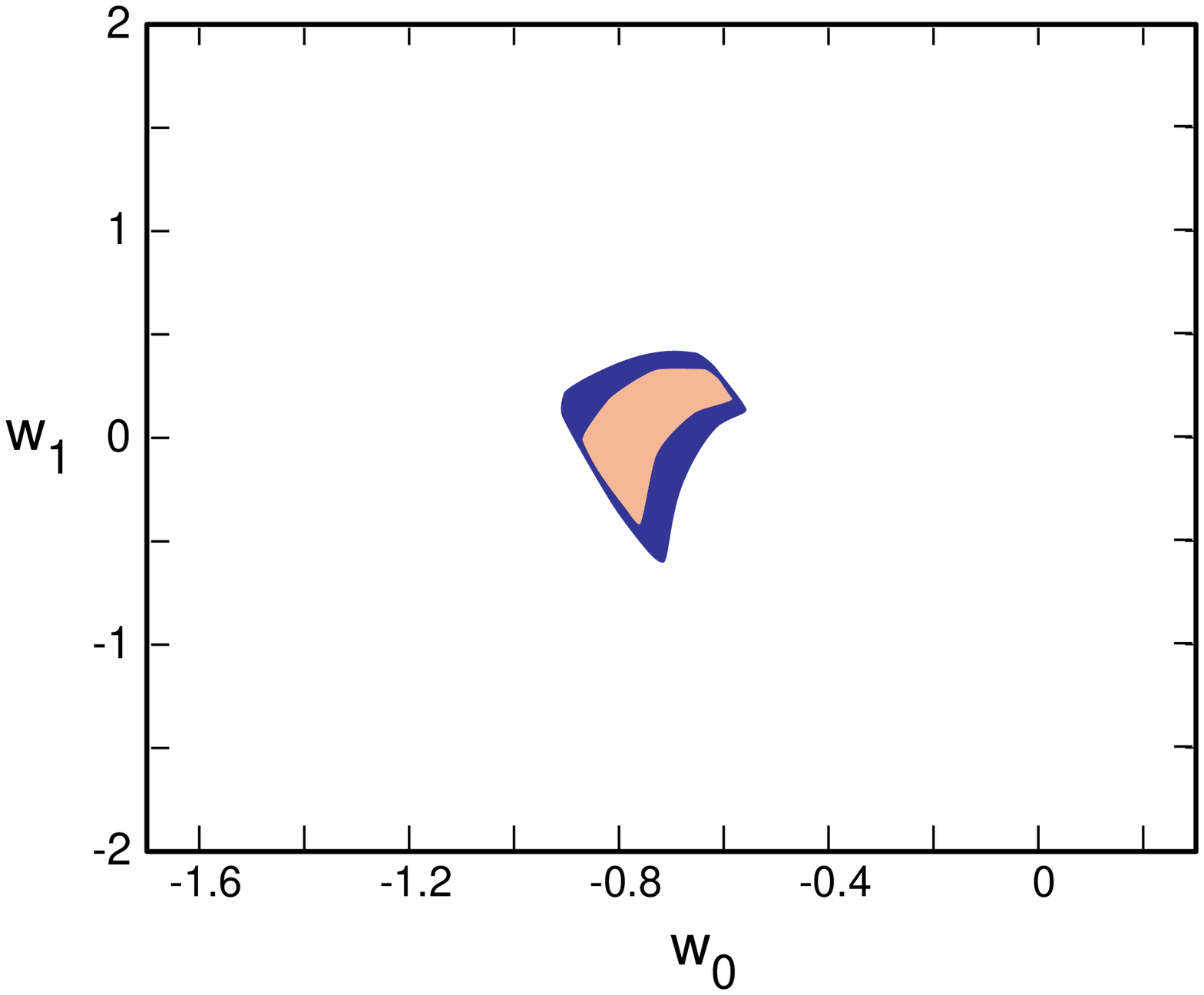} \
    \epsfxsize=50mm \epsfbox{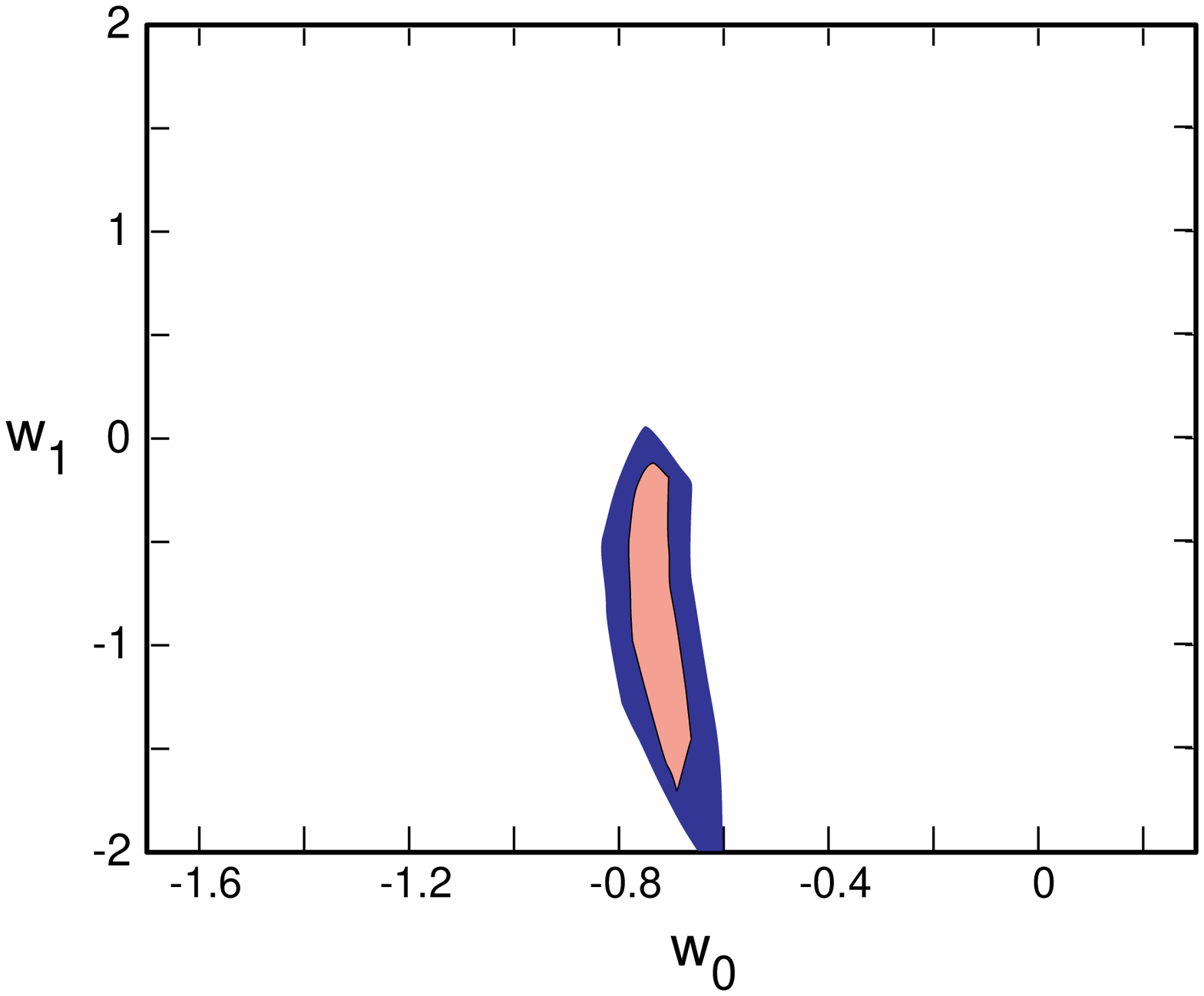} }
  \end{center}
  \caption{Likelihood contours (68\% (lighter) and 95\% (darker)
    C.L.) in the $(w_0,w_1)$
        plane, for fits to data generated
        from 3 different fiducial models.
        LEFT: $(w_0,w_1,\Omega_m)=(-0.7,0.8,0.3)$.
        MIDDLE: $(w_0,w_1,\Omega_m)=(-0.7,0.2,0.3)$.
        RIGHT: $(w_0,w_1,\Omega_m)=(-0.7,-0.8,0.3)$.
        Only the results shown in the left panel are
        inconsistent with a
        constant ($w_1=0$) $w_Q$ model.
       \label{der}}
\end{figure}

An EOS in which $w(z)$ has a large
positive time derivative is easier to detect than one which has a
large negative time derivative.  In either case, the derivative
must be large to be detected, as pointed out in Paper I,
 but here we are demonstrating that the challenge is
asymmetric. The point  is illustrated in Fig.~\ref{der}.  The
middle panel shows a fiducial model with a modest value of
$w_1=0.2$, and, as can be seen, this case cannot be distinguished
from a model in which $w_1=0$. That is, the 95\% CL contours
overlap the line $w_1=0$, corresponding to no time variation. The
left and right panels show cases in which $w_1=0.8$ and $-0.8$,
respectively.  The contours for the $w_1=0.8$ case (left) lie far
from the $w_1=0$ line, so the time variation is detectable in
this example.  On the other hand, the contours for the $w_1=-0.8$
case (right) overlay the $w_1=0$, so the time-variation is not
resolved.

This effect can be explained by considering the variation
of the total average EOS $w_T$ with respect to $w_1$,
$\Delta w_T=f(w_1)\Delta w_1$:
\begin{eqnarray}
  && f(w_1)=\Omega_Q\left[z+3w_Q\left(z-ln(1+z)\right)
        (1-\Omega_Q) \right].
        \label{erreq}
\end{eqnarray}
We consider $w_T$ because the measurements of $d_L$ are directly
sensitive to $w_T$, so that models can only be distinguished if
they have different $w_T$. As can be seen from eq.(\ref{erreq}),
$w_T$ is much less sensitive to changes in $w_1$ when it is
negative than when it is positive, mainly due to the value of
$\Omega_Q$ being larger for positive values of $w_1$, We conclude
that, in order to detect that $w_Q$ is time-dependent, it must be
that the time-variation is large, roughly $w_1 > 0.5$, and it
helps if $w_1$ is positive.  This corresponds to the case where
acceleration is becoming stronger as time evolves.

\section{Combining Supernovae with other Approaches}

Measurements to determine the EOS of the dark energy can be
direct or indirect. Direct methods, such as SNIa observations,
the Alcock-Paczynski (AP) test \cite{ap}, and  the cosmic
microwave background (CMB) attempt to measure the Hubble parameter
$H$, its derivative $H'$ and $\Omega_m$, or some function of them.
Indirect methods, such as structure formation aspects of the CMB
and measurements of large scale structure (LSS) try to infer
$w_Q(z)$ from its effects on structure evolution.

\begin{figure}
\vspace{-.75in}
\begin{center}
\epsfxsize=4.0 in \centerline{\epsfbox{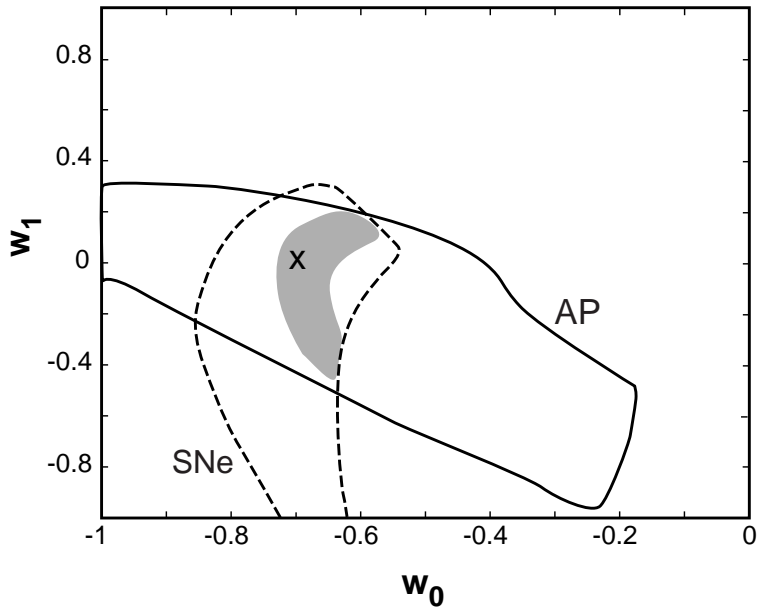}}
\end{center}
  \caption{ Two-sigma contours in the
 $(w_0, w_1)\equiv (w_Q(z=0), \, d w_Q/d z_0)$ plane
 for two idealized experiments. One
 measures thousands of supernovae between $z=0$ and
 $z=2$ (dashed contours). The supernovae are  divided into
 50 bins with a net  error of 1.4\% per bin.
 The  second  experiment is an optimistic estimate
 for the AP test (solid contours), assuming
  50 bins of Lyman-alpha clouds uniformly distributed
 between  $z=1.5$ and $z=3$ with each bin
 measured with an accuracy of 3\%.
 Both experiments assume a fiducial model with $\Omega_m=0.3$,
 $\Omega_Q=0.7$, $w_Q=-0.7 = const.$, indicated by the
 X. In both experiments $\Omega_m$ is marginalized over the
 range 0.2 to 0.4. The two-sigma joint likelihood for the two
 observations is shown in the shaded region.}  \label{contour}
  \end{figure}

An example of a complementary observation is the
the Alcock-Pazcynski (AP) test.
The physical transverse size of an object is given by
$d_T=d_A\Delta\theta=\frac{ r}{1+z}\Delta\theta$, $d_A$ being the
angular distance and $\Delta\theta$ the observed angular size. The
physical radial
size is $d_R=\int\sqrt{g_{rr}}dr=\frac{1}{(1+z)H(z)}\Delta z$.
For a population of spherical objects, the AP test is given
by equating the transverse and radial sizes:
$
  AP(z)=\frac{\Delta z}{\Delta \theta}=H(z)r(z)=
      H(z)\int_1^{1+z}\frac{dx}{H}.
$

The AP test on its own is not expected to improve the resolution
of the dark EOS since it has a more complex dependence on $w_Q$
than $d_L$. What does seem promising, as pointed out by McDonald
\cite{McDonald1,McDonald2}, is that the AP test can further
constrain the range of $\Omega_m$.

Fig.~\ref{contour} shows the likelihood contours assuming
optimistic anticipated errors over a continuous range between
$z=0$ and $z=2$ of 1.4\% for $d_L$, and, for the AP test,
50 bins between redshift $z=1.5$ and $z=3$ measured with
3\% error per bin.\cite{McDonald2}  
Both simulations represent 
highly optimistic assumptions about future measurements.
The results are interesting.  
The better constraint on $\Omega_m$ from the AP
test reduces the uncertainty in $w_0$, but does not 
significantly change the uncertainty in the time-variation,
 $w_1$.
This is not surprising since even a perfect
determination of $\Omega_m$ would leave a considerable
uncertainty in  $w_1$, as shown in Paper I.  To be sure,
the Alcock-Paczynski test is useful and worth pursuing, and 
a highly precise measurement combined with a highly precise
measurement of SNe could determine the present 
value of $w$ to within 15 or 20 percent.  However, it
does not help significantly with the particular problem of
pinning down the time-variation of the equation-of-state.

Measurements of the CMB anisotropy provide an additional probe of
$w(z)$.  This probe also suffers from a degeneracy problem, even
in the case where $w$ is constant.  The positions of the acoustic
peaks in the temperature anisotropy power spectrum depend on the
angular distance ($d_A$) to the last scattering surface which,
just like the luminosity distance for supernovae, depends on a
multi-integral over $w(z)$, $d_A=d_L/(1+z)^2$.  In addition, the
heights of the peaks depend on $\Omega_Q$ and the Hubble
parameter, $H_0=h\ 100\ {\rm km/parsec/sec}$.  When all effects
are considered, then, as shown by Huey, et al.  \cite{degen}, the
power spectrum is unchanged as certain combinations of $\Omega_m$,
$h$, and $w$ are varied. Consequently, none of these parameters
can be determined well by the CMB data alone.  Instead,
measurements can only constrain these parameters to a thin
two-dimensional surface
in this three-dimensional parameter subspace.

The reason why one might be optimistic about combining CMB
anisotropy and SN measurements is that the degeneracy surface for
the CMB anisotropy measurements is nearly orthogonal to the
degeneracy surface for the SN measurements  for the case of
constant $w$. Figure~8 illustrates the small overlap between the
SN and CMB degeneracy regions in the $\Omega_m$-$w$ plane. Other
authors have considered adding the CMB
contribution\cite{Huterer,Tegmark} but they have not included the
degeneracy aspect.  As we shall show below, introducing
time-varying $w(z)$ introduces additional degeneracy that spoils
the resolution even when the SN and CMB anisotropy measurements
are combined.

\begin{figure}
 \begin{center}
\vspace{-.5in}
  \epsfxsize=4 in \centerline{\epsfbox{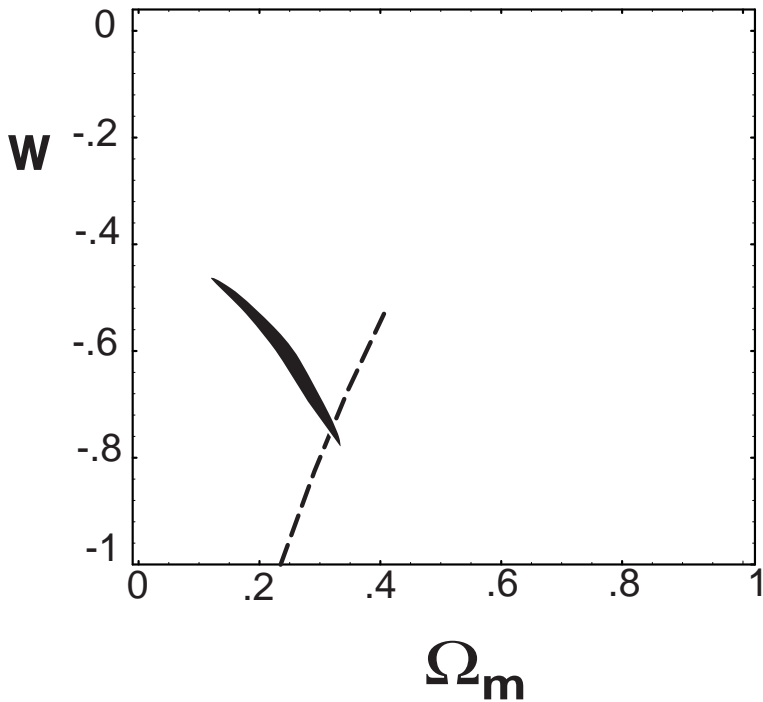}}
 \vspace{-.2in}
\end{center}
 \caption{
A simulation of the problem that  arises
if one assumes $w(z)$ is constant in the fitting procedure.
For a given fiducial model, the likelihood  fit for the
CMB anisotropy (dashed line) and SN luminosity distance-red
shift  (contour) observations  are illustrated.
The  degeneracy curve for the CMB assumes  cosmic
variance limited sampling, and the SN contour  assumes
1\% error in luminosity distance.  Each degeneracy region is
long and thin, and the two are nearly orthogonal.  Based
on the small overlap, one is tempted to conclude that
constancy of  $w$ is well established and its value is well
determined. However, that conclusion is absolutely wrong.
The fiducial model in this example actually
has a rapidly time-varying $w(z) = -2/3 - 1/6 z$
for $z<2$ and $w(z)=-1$ for $z>2$.
The degeneracy regions were computed assuming $w_1=0$, but,
if  $w_1$ is fixed at a value somewhat
less than  zero, say,  there are
once again two narrow
degeneracy regions which intersect over a small region, but the
 value of $w_0$ in the overlap region is significantly shifted.
 That is,
 the two experiments produce two degeneracy surfaces
 that intersect along a curve in the $w_1$ direction along
 which a degeneracy remains.
}
\end{figure}

Rather than do another complete survey, which is
a major technical challenge on its own,
we illustrate the degeneracy in parameter space
with a simple example in which
we consider the family of $w(z)$ of the form:
\begin{eqnarray}
w(z) & = &  w_0 + w_1 \,  z  \; \; \; {\rm for}  \;  z < 2
\nonumber   \\
& = &  w_0 +  2 \, w_1 \; \; \; {\rm for} \;  z \geq 2
\end{eqnarray}
This form was chosen to allow significant time variation recently
when $\Omega_{Q}$ is large and, in particular, to be similar to
the models considered in Paper I for $z<2$.  For $z<2$, the
degeneracy problem with respect to SN data was already
demonstrated and the $w_0$-$w_1$ degeneracy region was
characterized.  However, we could not simply maintain the linear
change in $w(z)$ with respect to $z$ out to the last scattering
surface at $z=1000$ because the value of $w$ would be ridiculously
non-physical. Hence, we cutoff the $z$-dependence at a value of
$z$ where $\Omega_Q$ is negligible and $w(z)$ is physically
plausible. We, then, maintain that condition back to the last
scattering surface.  For example, for $w=-2/3-1/6 z$ and
$(\Omega_Q)_0=.7$, at $z=2$ the dark energy contribution to the
total energy density is less than 15\%, which makes the details of
the $z$-dependence cutoff unimportant. From $z=2$ until last
scattering surface, this model will have $w=-1$.

The value of $w_0$ in our time-varying examples is fixed to be
$-2/3$ except where otherwise stated. In each of these models, we
also have $h=.65$, $\Omega_{Q}= .7$, $\Omega_{m} =.3$, and
$\Omega_{b}=.04$. Here $\Omega_b$ is the baryon density and
$\Omega_m$ is the total matter density (baryonic plus
non-baryonic).
 Note that  luminosity distance-red shift measurements are
not sensitive to $\Omega_b/\Omega_m$, but
the CMB measurements are.

The time-varying models were treated as the fiducial model, and
then a numerical search was performed for a constant EOS model
that is indistinguishable from the fiducial model based on the
combined measurement of the CMB and of supernovae. Models were
considered degenerate under the combined tests if: (1) the
percent difference between the luminosity distance-redshift
predictions for the two models is less than one percent out to
$z=2$ (the same criterion as in Paper I); and, (2) the CMB
predictions for the two models assuming a full-sky
cosmic-variance limited
measurement (no experimental error) cannot be distinguished to
better than 3$\sigma$. Both criteria are based on optimistic
predictions of what will be realistically possible.

\begin{figure}
 \begin{center}
\vspace{-.5in}
  \epsfxsize=5.0in \centerline{\epsfbox{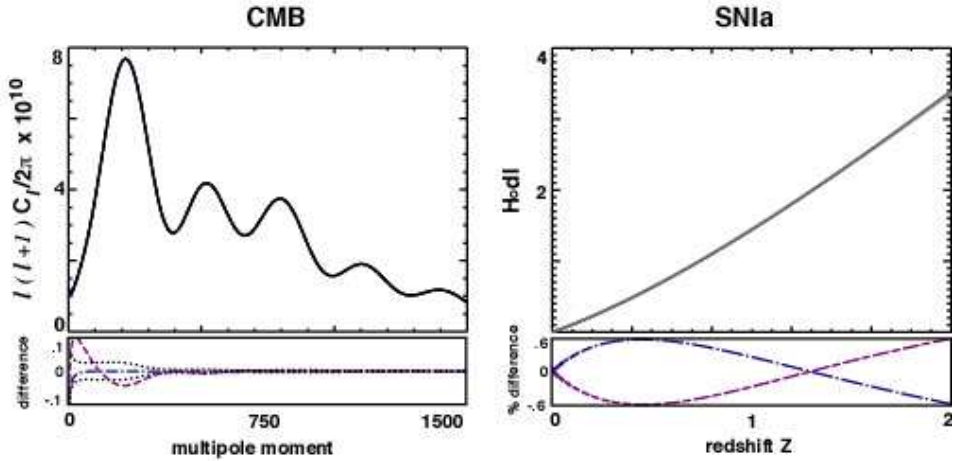}}
 \vspace{-.2in}
\end{center}
 \caption{
Illustration of the degeneracy problem for a model with
constant $w$ and two models with time-varying $w$ as discussed
in the text. The upper left hand panel compares the CMB
power spectra.  The lower left shows the differences between
the time-varying models and the constant $w$ model
and shows that they are less than or comparable
to the  full-sky cosmic variance theoretical uncertainty, the 
envelope shown in the figure (dotted lines).
The upper right panel  compares predictions for the luminosity
distance-red shift relation. The lower right panel shows the
differences with respect to constant model are less than
the 1\% resolution anticipated from supernovae measurements.
}
\end{figure}

For the CMB, distinguishability between a model with a constant
$w$ and a fiducial with a time-dependent $w$ was determined by a
log-likelihood analysis. The log-likelihood was calculated
according to the log-likelihood formula obtained by Huey, {\it et
al.}\cite{degen};
\begin{equation}
\L_{F C}= -\sum_l(\ell+ \frac{1}{2})\times
\left(1-\frac{C_{\ell}^{(F)}}{C_{\ell}^{(C)}}
+ \log\frac{C_{\ell}^{(F)}}{C_{\ell}^{(C)}} \right).
\end{equation}
The coefficients $C_{\ell}^{(F)}$ and $C_{\ell}^{(C)}$ are the
CMB multiple moments corresponding
to the fiducial and constant equation of state models, respectively.

Fig.~8 illustrates the problem that arises if one assumes $w(z)$
is constant in the fitting procedure. We have already observed
that this distorts results for the case of SN data alone.  Here
we show that the problem remains if CMB data is co-added.
Assuming $w$ is constant ($w_1 =0$), both measurements produce a
thin degeneracy region in the $\Omega_m$-$w$ plane. Based on the
small overlap, one is tempted to conclude that constancy of  $w$
is well established and its value is well determined. However,
this conclusion is absolutely wrong. In this example, the fiducial
model actually has a rapidly time-varying EOS $w(z) = -2/3 - 1/6
z$ for $z<2$ (which produces a change in $w$ of 50\% over this
range), and $w(z)=-1$ for $z>2$. The degeneracy regions were
computed assuming $w_1=0$. If $w_1$ were to be fixed at a
different value, once again the two measurements will give two
narrow degeneracy regions with a small overlap, but the value of
$w_0$ in this overlap region is significantly shifted. For
example, as shown in Fig~8, fixing $w_1=0$ results in $w_0=-.74$
with a few percent error. However, if $w_1$ were to be fixed at
its correct value $w_1=-1/6$, the result would have been
$w_0=-2/3$, again with a few percent error. But the central
values differ by more than 10\%. It is clear, then, that the two
measurements produce two degeneracy surfaces which intersect along
a degenerate curve which passes through a range of models with
varying values for $w_0$ and $w_1$, that remain degenerate under
the combined observations. For more complicated functional forms
of $w_Q(z)$ the degeneracy curve becomes a more complicated
higher dimensional surface, and the range of degeneracy in
parameter space (say, for $w_0$) increases.

The complications in the process of extracting the EOS from
both measurements are further illustrated in Fig.~9. There we show two
time-varying models with slopes $|w_1| > 0.1 $, one of which is
degenerate (by the log-likelihood test)  with a constant $w$ model with
$w=w_0=-.72$ and $\Omega_{b}=.04$, $\Omega_m=.27$, $\Omega_{Q}=.69$,
and $h=.64$.  The other can be barely resolved making the most
optimistic estimates about cosmic variance. A slight decrease in $w_1$,
or a slight decrease in experimental sensitivity would render the
second model degenerate. The lower two plots magnify the differences
between the predictions of the models. For the case of the CMB, we have
also shown the envelope based on the constant $w$ model corresponding
to the full-sky cosmic variance limit.  For the SN, we have constrained
the limits to lie between $\pm 1$\%.

\begin{figure}
 \begin{center}
\vspace{-.5in}
  \epsfxsize=4in \centerline{\epsfbox{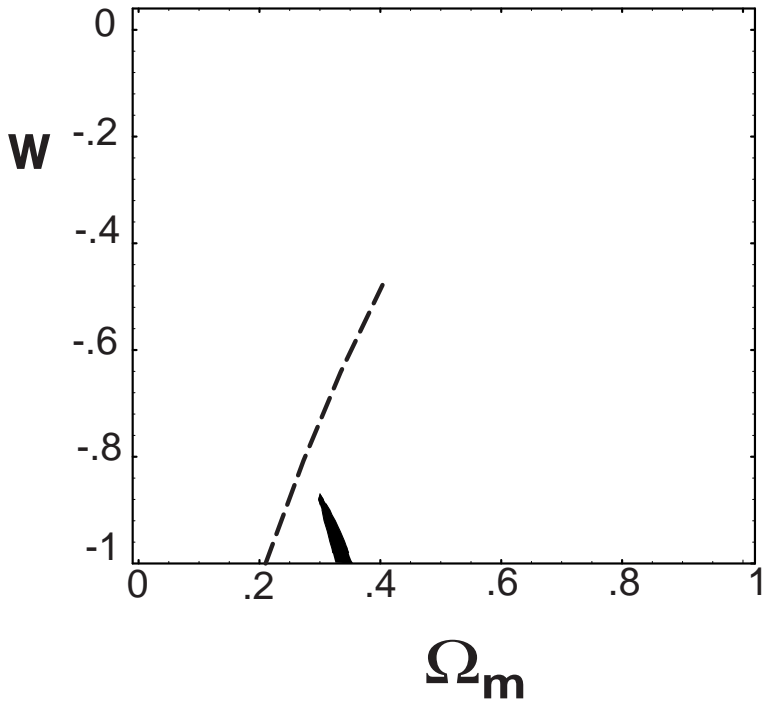}}
 \vspace{-.2in}
\end{center}
 \caption{
 The same as Fig.~8 but with a fiducial model with $w_1=1/3$.
 For cases like this
 with very rapid time-variation in $w$, a symptom is that the
 CMB and SN degeneracy regions do not overlap.  For
 $w_1$ large and positive, as in this example
 the SN contour (solid black) lies to the right  of the very thin
 CMB degeneracy region (dashed curve).
 For $w_1$ large and negative, the SN contour lies to the left.
}
\end{figure}

If $|w_1|$ is larger than 1/6
for our particular form of $w(z)$, we find that there is no overlap
between the degeneracy curve picked out by CMB measurements and
the degeneracy contour picked out by SN measurements
(where both fits assuming $w$ is constant).
An example is shown in Fig.~10.
In the case of negative (positive) $w_1$ the CMB measurements
that fit best suggest low (high) $\Omega_m$
whereas the SN measurements suggest high (low) $\Omega_m$.
If this absence of overlap were to be found in the real data,
an interpretation to pursue is that $w$ is rapidly time-varying.
Yet such an extreme scenario is not favoured by most theoretical
models, most of which predict a moderately time-varying $w$.
For the more likely case, in which the two measurements do overlap,
combining them reduces degeneracy by only a modest amount,
generally not even enough to decide whether the dark energy
in the universe has a time varying equation of state or not.

Co-adding the CMB to the SN data represents an improvement in the
sense that $w_1$ and $w_0$ are more constrained than with SN data
alone based on our earlier analysis or in Paper I. The
improvement is by a factor of four or so {\it assuming a linear
form for $w(z)$}, which is
significant.  However, there remains a large
uncertainty in the EOS. Furthermore,
we would  stress once
again that the accuracy in determining  $w$ strongly depends on
its assumed functional form.
The range of degeneracy
obtained for $w_1$  (a bit more than $\pm 0.1$)
in our example underestimates the degeneracy
 for general $w(z)$.
For example, for parabolic forms, the uncertainty in $w_1$ blows
up to $\pm 0.5$.
Given the extraordinarily precise data that has
been brought to bear, the allowed variation in $w_0$, $w_1$ and
$\Omega_m$  is disappointing.

\section{Conclusions}
\label{conc}

An important challenge for observational cosmology is to measure
the equation of state of the dark energy, $w_Q(z)$. This can
provide important information about the fundamental physics that
is responsible for the accelerated expansion of the universe.
Measurements of the distance-red shift relation using supernovae,
perhaps combined with other direct methods such as the
Alcock-Paczyinski test or the cosmic microwave background, would
appear to be promising methods. Indeed, analyses based on the
{\it a priori} assumption that $w_Q(z)$ is constant  suggest that
$w_Q$ can be resolved to 5\% accuracy or better.

In this paper and Paper I, though, we have uncovered a number of
problems and pitfalls that arise when trying to determine
$w_Q(z)$ without making prior assumptions. Our lessons may be
summarized as follows:
\begin{itemize}
\item Because measures of luminosity or angular distance depend
on integrals over $w_Q(z)$, a first degeneracy problem arises in which
neither the current value and nor its time-variation can be
resolved to any useful accuracy. (Sec.~II)
\item Since the effect of dark energy on the luminosity distance depends on
the combination $w_Q  \Omega_Q$ rather than $w_Q$ itself, a
second degeneracy problem arises in which $w_Q$ and $\Omega_Q$
are changed simultaneously so as to keep $w_Q  \Omega_Q$ fixed.
(Sec.~III)
\item  Although SN measurements may extend to $z=2$, they are most
sensitive to the behavior of $w_Q(z)$ at a modest value of $z^*
\approx 0.1-0.4$. (Sec.~III)
\item Consequently,  if there were only  the first degeneracy problem,
$w_Q(z)$ could be well-resolved at $z=z^*$ even though it is not
well-resolved for other values of $z$. Unfortunately, the
resolution of $w_Q(z^*)$ is totally degraded when one includes
uncertainty in $\Omega_Q$ and the second degeneracy problem.
(Sec.~III, especially Fig. 2)
\item The common practice of fitting data assuming that $w_Q(z)$
is constant can lead to grossly distorted results. Similarly,
the common practice of assuming $w_Q \ge -1$
can lead to grossly distorted results.
Fig.~5 shows a dramatic example  in which these practices lead
to the conclusion that
$w_Q=-1$ and  is well-resolved when, in reality,  $w_Q>-1$
and rapidly increasing. (Sec. IV)
\item Time-variation of $w_Q$ is more easily detected if
$w_Q(z)$ is an increasing function of $z$ rather than decreasing.
(Sec. V)
\item To resolve $w_Q(z)$ with supernova data, an additional
test is needed.
Given optimistic estimates of experimental
uncertainties, the Alcock-Paczynski test combined with the 
supernovae measurements can constraint the current value of $w$ to
within 20 percent or so. However, Neither the Alcock-Paczynski test
nor microwave background
anisotropy measurements provide  the needed resolution to constrain 
the time-variation.
(Sec. VI)
\end{itemize}

Our principal conclusion is that a new  test is required to
achieve the goal of measuring $w_Q(z)$. In devising a new test,
the two considerations must be precision and model dependence.
Thus far, among the measurements that we have considered, the
measurements which are precise give constraints on  $w_Q$ that are
highly model dependent, leading to degeneracy problems. Tests
which are not model dependent turn out to be difficult to measure
precisely.  So, there lies the challenge.

In considering alternatives, it is critical to include practical
estimates of their uncertainties.
Furthermore, one must consider how the new tests, themselves,
depend on $w_Q(z)$. For example, claims have been made that
$\Omega_m$ and $\Omega_Q$ have been or will be measured very
accurately by measurements of the cosmic microwave background
\cite{Turner:2001mw}.  However, those estimates are based on
assuming that $w_Q=-1$.  Making no prior assumption about
$w_Q(z)$, a degeneracy problem once again arises\cite{degen} that
spoils the resolution of $\Omega_Q$ and $w_Q(z)$, as discussed in
Sec. VI.

While trying to devise a new test to determine $w_Q$, it is worth
mentioning that a precise measurement of $H'$ will be extremely
useful.
The dependence of $w_Q$ on $H$, $H'$ (prime denotes a derivative
with respect to $x=1+z$) and $\Omega_m$ is given by
$   w_Q=\frac{\frac{2}{3} x HH'-H^2}
    {H^2-\Omega_m H_0^2\ x^3}.$
A good measurement of $H'$ is clearly crucial to the resolution
of $w_Q$, but current tests do not probe $H'$ directly.
The next best option is to measure $H(z)$, and then estimate $H'$
by calculating its derivative. Obviously, this worsens the
resolution for $H'$ and increases the uncertainty in $w_Q$.

Three additional approaches that we have not tried yet are
measuring the time dependence of
structure growth on $w(z)$;\cite{Weller3} gravitational lensing; 
and direct measurements of $dz/dt$
(to be discussed elsewhere\cite{Loeb}).

\section{Acknowledgments}


We thank A. Albrecht, G. Efstathiou, D. Eichler,
P. McDonald, and B. Paczynski for helpful comments, and D.  Oaknin for
valuable programming assistance. This research was supported by grant
No.~1999071 from the United States-Israel Binational Science Foundation
(BSF) (IM and RB) and by the US Department of Energy grant
DE-FG02-91ER40671 (JM and PJS).



\begin{thebibliography}{}

\bibitem{SCP}
S. Perlmutter, {\it et al.}, {\it Ap. J.}  {\bf 517}, 565 (1999).

\bibitem{HZS}
A.G. Riess, {\it et al.}, {\it Ap. J.}  {\bf 116}, 1009 (1998).

\bibitem{EOS}
S. Perlmutter, M.~S. Turner and M. White, {\it Phys. Rev. Lett.}
{\bf  83},  670
(1999);  P. Garnavich, {\it et al.}, {\it Ap. J.}  {\bf 509} 74 (1998).


\bibitem{OS}
See, for example,
J. P. Ostriker and P.J. Steinhardt, {\it Nature} {\bf 377}, 600 (1995);
L.M. Krauss and M.S. Turner,
{\it Gen. Rel. Grav.}{\bf  27}, 1137 (1995).

\bibitem{Bahcall}  See, for example, N. Bahcall, J.P. Ostriker. S.
Perlmutter, and P.J. Steinhardt, {\it Science} {\bf 284}, 1481-1488,
(1999)
and references therein.

\bibitem{CDS}  R.R. Caldwell, R. Dave and P.J. Steinhardt, {\it Phys.
Rev. Lett.} {\bf 80}, 1582 (1998).

\bibitem{Maor:2001jy}
I.~Maor, R.~Brustein and P.~J.~Steinhardt,
Phys.\ Rev.\ Lett.\  {\bf 86}, 6 (2001).


\bibitem{other}
N. Weiss, {\it Phys. Lett. B} {\bf 197}, 42 (1987);
B. Ratra and J.P.E. Peebles,  {\it Ap. J.} , 325, L17
(1988);
C.Wetterich, {\it Nucl. Phys. B}{\bf 302}, 668 (1988),
and  {\it Astron. Astrophys.} {\bf 301}, 32 (1995);
J.A. Frieman, {\it et al.}  {\it Phys. Rev. Lett.} {\bf 75}, 2
077 (1995);
K. Coble, S. Dodelson,  and J. Frieman, {\it Phys. Rev. D}
{\bf 55}, 1851 (1995);
P.G. Ferreira and M. Joyce, {\it  Phys. Rev. Lett.} {\bf 79},
4740 (1997);
{\it Phys. Rev. D} {\bf 58}, 023503 (1998);
 C. Armendariz-Picon, V. Mukhanov, and P.J. Steinhardt,
 {\it Phys. Rev. Lett.} {\bf 85},
 4438-41  (2000).

\bibitem{Efst}  G. Efstathiou, {\it MNRAS} 310, 842 (1999).

\bibitem{Podariu}
S.~Podariu, P.~Nugent and B.~Ratra,
Astrophys.\ J.\  {\bf 553}, 39 (2001).

\bibitem{astier}
P. Astier,
astro-ph/0008306.

\bibitem{Weller}
J.~Weller and A.~Albrecht,
Phys.\ Rev.\ Lett.\  {\bf 86}, 1939 (2001).

\bibitem{Chiba}
T.~Chiba and T.~Nakamura,
Phys.\ Rev.\ D {\bf 62}, 121301 (2000).

\bibitem{Barger}
V.~Barger and D.~Marfatia,
Phys.\ Lett.\ B {\bf 498}, 67 (2001).

\bibitem{Chevallier}
M.~Chevallier and D.~Polarski,
Int.\ J.\ Mod.\ Phys.\ D {\bf 10}, 213 (2001).

\bibitem{Goliath}
M.~Goliath, R.~Amanullah, P.~Astier, A.~Goobar and R.~Pain,
astro-ph/0104009.

\bibitem{trentham}
N. Trentham,
astro-ph/0105404.

\bibitem{gudmun}
E.~H.~ Gudmundsson and G. Bjornsson,
astro-ph/0105547.

\bibitem{weller2}
J. Weller and A. Albrecht,
astro-ph/0106079.

\bibitem{Ng}
S.~C.~Ng and D.~L.~Wiltshire,
Phys.\ Rev.\ D {\bf 64}, 123519 (2001).

\bibitem{Kujat}
Jens Kujat, Angela M.~ Linn, Robert J.~ Scherrer, David H.~
Weinberg,
astro-ph/0112221.

\bibitem{Hut} D. Huterer and M.S. Turner,
{\it Phys. Rev. D} {\bf  60}, 081301 (1999).

\bibitem{Staro}
T.~D.~Saini, S.~Raychaudhury, V.~Sahni and A.~A.~Starobinsky,
Phys.\ Rev.\ Lett.\  {\bf 85} (2000) 1162.


\bibitem{Boisseau}
B.~Boisseau, G.~Esposito-Farese, D.~Polarski and
A.~A.~Starobinsky,
Phys.\ Rev.\ Lett.\  {\bf 85}, 2236 (2000).

\bibitem{Wang}
Y.~Wang and P.~M.~Garnavich,
Astrophys.\ J.\  {\bf 552}, 445 (2001).


\bibitem{Huterer}
D.~Huterer and M.~S.~Turner,
Phys.\ Rev.\ D {\bf 64}, 123527 (2001).


\bibitem{Tegmark}
M.~Tegmark,
astro-ph/0101354.


\bibitem{Corasaniti}
P.~S.~Corasaniti and E.~J.~Copeland,
astro-ph/0107378.

\bibitem{Lovelace}
Y.~Wang and G.~Lovelace,
Astrophys.\ J.\  {\bf 562}, L115 (2001).


\bibitem{SNAP} An example is the proposed SNAP
(Supernova  Acceleration Probe) satellite, http://snap.lbl.gov

\bibitem{ap}
C. Alcock and B. Paczynski, Nature {\bf 281}, 358 (1979).

\bibitem{McDonald1}
P.~McDonald and J.~Miralda-Escude,
Astrophys.\ J.\  {\bf 518},24 (1999).

\bibitem{McDonald2} P. McDonald, astro-ph/0108064.


\bibitem{degen}   G. Huey, L. Wang, R. Dave,  R. R. Caldwell and
P.J. Steinhardt, {\it Phys. Rev. D}{\bf 59}, 063005  (1999).

\bibitem{Turner:2001mw} M.S. Turner, astro-ph/0106035.

\bibitem{Weller3} J. Weller, R. Battye, and R. Kneissl,
astro-ph/0110353.

\bibitem{Loeb}  D. Wesley, A. Loeb, and P.J. Steinhardt,
to appear.

\end{thebibliography}
\end{document}